\documentclass[11pt,draftclsnofoot,onecolumn]{IEEEtran}
\usepackage{fancyhdr}
\usepackage{epsfig,graphics,subfigure,amsthm,psfrag,amsmath,amssymb,cite,citesort}
\usepackage{algorithm}               
\usepackage{algorithmic}             
\usepackage{xcolor}
\usepackage[center]{caption}
\allowdisplaybreaks

\newtheorem{Definition}{Definition}

\renewcommand{\Bmatrix}[1]{\begin{bmatrix}#1\end{bmatrix}}

\newcommand{\trace}[1][]{\ifthenelse{\isempty{#1}}{\text{tr}}{\text{tr}\left(#1\right)}}
\newcommand{\rank}[1][]{\ifthenelse{\isempty{#1}}{\text{rank}}{\text{rank}\left(#1\right)}}

\newcommand{\defeq}{\triangleq}

\newcommand{\E}[1][]{\ifthenelse{\isempty{#1}}{\mathbb{E}}{\mathbb{E}\left(#1\right)}}

\begin{document}
{\fontsize{10.5pt}{12.5pt}\selectfont

\title{Precoding Methods for MISO Broadcast Channel with Delayed CSIT}

\author{\IEEEauthorblockN{Xinping Yi, \emph{Student Member, IEEE}, and David Gesbert, \emph{Fellow, IEEE}}
\thanks{X.~Yi and D. Gesbert are with the Mobile Communications Dept., EURECOM, 06560 Sophia Antipolis, France (email: \{xinping.yi,david.gesbert\}@eurecom.fr).
Part of this work has been presented at ICASSP 2012 in Kyoto, Japan.}}

\maketitle 

\begin{abstract}
Recent information theoretic results suggest that precoding on the multi-user downlink MIMO channel with delayed channel state information at the transmitter (CSIT) could lead to data rates much beyond the ones obtained without any CSIT, even in extreme situations when the delayed channel feedback is made totally obsolete by a feedback delay exceeding the channel coherence time. This surprising result is based on the ideas of interference repetition and alignment which allow the receivers to reconstruct information symbols which canceling out the interference completely, making it an optimal scheme in the infinite SNR regime. In this paper, we formulate a similar problem, yet at finite SNR.  We propose a first construction for the precoder which matches the previous results at infinite SNR yet reaches a useful trade-off between interference alignment and signal enhancement at finite SNR, allowing for significant performance improvements in practical settings. We present two general precoding methods with arbitrary number of users by means of virtual MMSE and mutual information optimization, achieving good compromise between signal enhancement and interference alignment. Simulation results show substantial improvement due to the compromise between those two aspects.
\end{abstract}


\IEEEpeerreviewmaketitle

\section{Introduction}
Multi-user MIMO systems (or their information-theoretic counterparts ``MIMO broadcast channels"),  have recently attracted considerable attention from the research community and industry alike. Success is due to their ability to enhance the wireless spectrum efficiency by a factor equal to the number $N$ of antennas installed at the base station, with little restriction imposed on the richness of the multipath channel, the presence or absence of a strong line of sight channel component, and the fact it can easily accommodate single antenna mobile devices. On the downlink of such systems, the ability to beamform (i.e. linearly precode) multiple data streams simultaneously to several users (up to $N$) comes nevertheless at a price in terms of requiring the base station transmitter to be informed of the channel coefficients of all served users \cite{Gesbert:07}. In frequency division duplex scenarios (the bulk of available wireless standards today), this implies establishing a feedback link from the mobiles to the base station which can carry CSI related information, in quantized format. A common limitation of such an approach, perceived by many to be a key hurdle toward a more widespread use of MU-MIMO methods in real-life networks, lies in the fact that the feedback information typically arrives back to the transmitter with a delay which may cause a severe degradation when comparing the obtained feedback CSIT with the actual current channel state information. Pushed to the extreme, and considering a feedback delay with the same order of magnitude as the coherence period of the channel, the available CSIT feedback becomes completely obsolete (uncorrelated with the current true channel information) and, seemingly non exploitable in view of designing the precoding coefficients.

Recently, this commonly accepted viewpoint was challenged by an interesting information-theoretic work which established the usefulness of stale channel state information in designing precoders achieving significantly better rate performance than  what is obtained without  any CSIT \cite{Maddah-Ali:10}. The premise in~\cite{Maddah-Ali:10} is a time-slotted MIMO broadcast channel with a common transmitter serving multiple users  and having  a delayed version of the correct CSIT, where the delay causes the CSIT to be fully uncorrelated with the current channel vector information. In this situation, it is shown that the transmitter can still exploit the stale channel information: The transmitter tries to reproduce the interference generated to the users in the previous time slots, a strategy we refer in this paper as interference repetition, while at the same time making sure the forwarded interference occupies a subspace of limited dimension, compatible with its cancelation at the user's side, a method commonly referred to as interference alignment \cite{Cadambe:08,Maddah-Ali:08}. Building on such ideas, \cite{Maddah-Ali:10} constructs a transmission protocol referred as the MAT protocol which was shown to achieve the  maximum Degrees-of-Freedom (DoF) for the delayed CSIT broadcast MIMO channel. Precoding on delayed CSIT MIMO channels have recently attracted more interesting work, dealing with DoF analysis on extended channels, like the X channel and interference channels~\cite{Maleki:10,Ghasemi:11,Vaze:11}, but also performance analysis including effects of feedback~\cite{Xu:11} and training~\cite{Adhikary:11}. The DoF is a popular information theoretic performance metric indicating the number of interference-free simultaneous data streams which can be communicated over this delayed CSIT channel at infinite SNR, also coinciding with the notion of pre-log factor in the channel capacity expression. In the example of the two antenna transmitter, two user channel, the maximum DoF was shown in~\cite{Maddah-Ali:10} to be $\frac{4}{3}$, less than the value of 2 which would be obtained with perfect CSIT, but strictly larger than the single DoF obtained in the absence of any CSIT. This means that completely obsolete channel feedback is actually useful.

Although fascinating from a conceptual point of view, these results are intrinsically focussed on the asymptotic SNR behavior, leaving aside in particular the question of how shall precoding be done practically using stale CSIT at finite SNR. This paper precisely tackles this question. In what follows we obtain the following key results:
\begin{itemize}
\item We show finite SNR precoding using delayed CSIT can be achieved using  a combination of interference repetition, alignment together with a signal enhancement strategy.
\item We propose a precoder construction generalizing the ideas of \cite{Maddah-Ali:10}, namely Generalized MAT (GMAT), where a compromise between interference alignment and orthogonality within the desired signal channel matrix is striken, and generalize it to the scenario with arbitrary number of users.
\item The precoder coefficients are interpreted as beamforming vector coefficients in equivalent interference channel scenario, which can be optimized in a number of ways, including using an MMSE metric, and mutual information metric. {To our best knowledge, the optimization of a finite SNR precoding scheme based on delayed feedback has not yet been addressed.}
\end{itemize}
Numerical evaluation reveal a substantial performance benefit in terms of data rate in the low to moderate SNR region, but coinciding with the performance of \cite{Maddah-Ali:10} when the SNR grows to infinity. Note that a preliminary set of results were reported recently in~\cite{YiG} for the 2-user case, while this paper provides a generalization to the case of arbitrary number of users.

The rest of the paper is organized as follows. In Section II, the channel model of interest is described and the proposed GMAT protocol is detailed first in the 2-user case then is generalized to the $K$-user case. Section III focuses on the precoder optimization method based on MMSE and mutual information criteria. Discussion on the multiplexing gain and an interesting interpretation from an equivalent MIMO interference channel is given in Section IV.  Numerical examples showing the advantages of the new methods are discussed in section V. Finally, Section VI concludes the paper.

\textbf{Notation}:
Matrices and vectors are represented as uppercase and lowercase letters, and transpose and conjugate transpose of a matrix are denoted as $(\cdot)^T$ and $(\cdot)^H$, respectively. Further, $\text{Tr}(\cdot)$, $\|\cdot\|$ and $\|\cdot\|_F$ represent the trace of a matrix, the norm of a vector and a Frobenius norm of a matrix. We reserve $[\mathbf A]_{m,n}$ to denote the element at the $m$-th row and $n$-th column of matrix $\mathbf A$, and $|\mathcal{S}|$ to the cardinality of the set $\mathcal{S}$. Finally, an order-$k$ message denoted by $u_{\mathcal{S}}$ ($|\mathcal{S}|=k$) refers to a linear combination of $k$ distinct symbol vectors intended to $k$ different users in set $\mathcal{S}$.

\section{System Model}
Consider a $K$-user MU-MIMO downlink system with a transmitter equipped with $K$ antennas and $K$ single-antenna users. A time slotted transmission protocol in the downlink direction is considered, where the multi-antenna channel vector from the transmitter to $i$-th user, in the $j$-th time slot, is denoted by ${\bf h}_i^T(j)=[h_{i1}(j) \ \cdots \  h_{iK}(j)]$. We denote by $\mathbf x(j)$ the $K \times 1$ vector of signals sent from the array of $K$ transmit antennas. As in~\cite{Maddah-Ali:10}, the point made in this paper is that delayed feedback can be of use to the transmitter including the extreme situation where a feedback delay of one unit of time creates a full decorrelation with the current downlink channel. For this reason, we base ourselves on the framework of so-called delayed CSIT~\cite{Maddah-Ali:10,Maleki:10,Ghasemi:11,Vaze:11,Xu:11,Adhikary:11} by which at time $j$, it is assumed that user-$i$ has perfect knowledge of $\{{\bf h}_i(t)\}_{t=1}^{j}$ and of the delayed CSIT of other users $\{{\bf h}_k(t)\}_{t=1}^{j-1},k \ne i$, while the transmitter are informed perfectly $\{{\bf h}_i(t)\}_{t=1}^{j-1}, \forall i$. Furthermore, we make no assumption about any correlation between the channel vectors across multiple time slots (could be fully uncorrelated), making it is impossible for the transmitter to use classical MU-MIMO precoding to serve the users, since the transmitter possesses some CSIT possibly independent from the actual channel.

Recently, Maddah-Ali and Tse~\cite{Maddah-Ali:10} proposed an algorithm under such delayed CSIT setting obtaining DoF strictly beyond that obtained without any CSIT, even in extreme situations when the delayed CSIT is made totally obsolete. The key ideas lie in interference repetition and alignment. Doing so, the users are able to reconstruct the signals overheard in previous slots to allow them to cancel out the interference completely. Particularly, in the 2-user case, it is assumed that three time slots are used to send a total of four symbols (two for each user), yielding an average rate efficiency of $4/3$ symbols/channel use, while in the 3-user case, it delivers total 18 symbols in 11 time slots, providing $\frac{18}{11}$ DoFs. Generally speaking, when there are $K$ users, a $K$-phase transmission protocol is proposed achieving the maximum DoF $\frac{K}{1+\frac{1}{2}+\cdots+\frac{1}{K}}$.
Although such rates are inferior to the ones obtained under the full CSIT setting ($K$ symbols/channel use for $K$ antenna system), they are substantially higher than what was previously reported for the no CSIT case (1 symbol/channel use regardless of $K$).

Although optimal in terms of the DoF, at infinite SNR, we point out that the above approach can be substantially improved at finite SNR. The key reason is that, at finite SNR, a good scheme will not attempt to use all DoFs to eliminate the interference but will try to strike a compromise between interference canceling and enhancing the detectability of the desired signal in the presence of noise. Taking into account this property of basic receivers leads us to revisit the design of the protocol and in particular the design of the precoding coefficients as function of the knowledge of past channel vectors under the name of GMAT. 

First, we proceed by reviewing the proposed protocol in the 2-user case, highlighting the connections with the original MAT algorithm. We then generalize the protocol to respectively the 3 and $K$-user cases. In the next section, we then turn to the problem of the optimization of the precoders.

\subsection{GMAT for the 2-user Case}
Here, we introduce the concept of the GMAT algorithm in the 2-user case.
Note that the transmission in the first two time slots is identical to the MAT algorithm, with
\begin{eqnarray}
\mathbf x(1)=\mathbf s_A,~~~ \mathbf x(2)=\mathbf s_B
\end{eqnarray}
where $\mathbf x(t)$ ($t=1,2$) is the $2 \times 1$ signal vector sent from the transmitter at time slot $t$, ${\bf s}_A$ and ${\bf s}_B$ are $2 \times 1$ symbol vectors intended to user A and B, respectively, satisfying $\mathbb{E}\{\mathbf s_i \mathbf s_i^H\}=\mathbf I$. In the third time slot, the transmitter now sends
\begin{eqnarray}
\mathbf x(3)=\begin{bmatrix} u_{AB} \\ 0 \end{bmatrix}
\end{eqnarray}
where $u_{AB}$ corresponds to an order-2 message (i.e., a combination of two individual user messages in the following form)
\begin{eqnarray}
u_{AB}= {\bf w}_1^T {\bf s}_A + {\bf w}_2^T {\bf s}_B \label{eq:msg-realization-2-user}
\end{eqnarray}
where $\mathbf w_1$ and $\mathbf w_2$ are precoding vectors satisfying the power constraint $\|\mathbf w_1\|^2+\|\mathbf w_2\|^2 \le 2$ and can be a function of $\mathbf h_i(1)$ and $\mathbf h_i(2)$ according to the delayed CSIT model. Note that this power constraint balances the transmit power used over three time slots. The signal vector received over the three time slots at user A is given by:
\begin{eqnarray} \label{eq:chennel-model}
\bar{\mathbf y}_A &=& \sqrt{\frac{P}{2}} \bar{\mathbf H}_{A1} \mathbf s_A + \sqrt{\frac{P}{2}} \bar{\mathbf H}_{A2} \mathbf s_B + \mathbf n_A,
\end{eqnarray}
where $\bar{\mathbf y}_A=[y_A(1)\ y_A(2)\ y_A(3)]^T$ is the concatenated received signal vector at user A in overall three time slots, $\mathbf n_A = [n_A(1)\ n_A(2)\ n_A(3)]^T$ is the Gaussian noise vector with zero-mean and unit-variance, $P$ is the total transmit power in each time slot, and the effective signal and interference channel matrices are
\begin{eqnarray} \label{eq:channel-vector-GMAT-A}
  \bar{\mathbf H}_{A1}=\begin{bmatrix} \mathbf h_A^T(1) \\ \mathbf 0 \\ h_{A1}(3) \mathbf w_1^T \end{bmatrix} ,\  \bar{\mathbf H}_{A2}=\begin{bmatrix} \mathbf 0 \\ \mathbf h_A^T(2) \\ h_{A1}(3) \mathbf w_2^T \end{bmatrix},
\end{eqnarray}
and, by analogy, for user B, we get
\begin{eqnarray}
\bar{\mathbf y}_B &=& \sqrt{\frac{P}{2}} \bar{\mathbf H}_{B1} \mathbf s_A + \sqrt{\frac{P}{2}} \bar{\mathbf H}_{B2} \mathbf s_B + \mathbf n_B,
\end{eqnarray}
where the interference and signal matrices are:
\begin{eqnarray} \label{eq:channel-vector-GMAT-B}
  \bar{\mathbf H}_{B1}=\begin{bmatrix} \mathbf h_B^T(1) \\ \mathbf 0 \\ h_{B1}(3) \mathbf w_1^T \end{bmatrix} ,\  \bar{\mathbf H}_{B2}=\begin{bmatrix} \mathbf 0 \\ \mathbf h_B^T(2) \\ h_{B1}(3) \mathbf w_2^T \end{bmatrix}.
\end{eqnarray}

\subsubsection{A Particular Case (MAT Algorithm)}
We point out that the MAT algorithm~\cite{Maddah-Ali:10} can be derived as a particular case of the above method, with ${\bf w}_1$ and ${\bf w}_2$ specified as
\begin{eqnarray} \label{eq:MAT-w}
{\mathbf w}_1 = \mathbf h_B(1),~~~~{\mathbf w}_2 = \mathbf h_A(2).
\end{eqnarray}
The key idea behind the original MAT solution in (\ref{eq:MAT-w}) is that the interference ${\bf s}_B$ seen by user A arrives with an effective channel matrix $\bar{\mathbf H}_{A2}$ which is of rank one, making it possible for user A to combine the three received signals in order to retrieve ${\bf s}_A$ while canceling out ${\bf s}_B$ completely. This process is referred to as alignment of interference signal ${\bf s}_B$, as it mimics the approach taken in interference channels in e.g.~\cite{Cadambe:08}. A similar property is exploited in (\ref{eq:MAT-w}) at user B as well by making $\bar{\mathbf H}_{B1}$ be rank 1.
\subsubsection{Interpretation of GMAT v.s.~MAT}
A drawback of the original MAT solution in (\ref{eq:MAT-w}) is to optimize the precoders from the point of view of interference alone while the signal matrices $\bar{\mathbf H}_{A1}$ and $\bar{\mathbf H}_{B2}$ are ignored. Although this approach is optimal from an information theoretic (multiplexing gain) point of view, it is suboptimal at finite SNR.

In contrast, here, the role of introduced beamformer ${\bf w}_1$ is to strike a balance between aligning the interference channel of ${\bf s}_A$ at user B and enhancing the detectability of ${\bf s}_A$ at user A. In algebraic terms this can be interpreted as having a compromise between obtaining a rank deficient $\bar{\bf H}_{B1}$ and an orthogonal matrix for $\bar{\bf H}_{A1}$. When it comes to ${\bf w}_2$, the compromise is between obtaining a rank deficient $\bar{\bf H}_{A2}$ and an orthogonal matrix for $\bar{\bf H}_{B2}$. How to achieve this trade-off in practice is addressed in Section III. Meanwhile, we show how the above transmission protocol can be extended to the 3-user and then the $K>3$ user cases.

{It is also important to note there might be alternative fashions of constructing finite SNR precoders based on delayed CSIT. For instance, an interesting question is: Can delayed feedback be exploited already in the second time slot with gains on the finite SNR performance? The intuitive answer to this question is yes. However, the use of precoders in the last time slot only generates a strong symmetry and handling of the users, which in turn allows for closed-form and insightful solutions. This symmetric property is also maintained in the MAT algorithm.}

\subsection{GMAT for the 3-user Case}
Similarly to the MAT algorithm, the proposed GMAT sends 18 symbols in a total of three phases, which include 6, 3, and 2 time slots, respectively, giving an effective rate of $\frac{18}{11}$ symbols/slot. In the first phase, 6 symbol vectors carrying all 18 symbols are sent in 6 consecutive time slots in a way identical to the initial MAT
\begin{eqnarray}
  \mathbf x(1) = \mathbf s_A^1,\ \mathbf x(2) = \mathbf s_B^1,\ \mathbf x(3) = \mathbf s_C^1,
  \mathbf x(4) = \mathbf s_A^2,\ \mathbf x(5) = \mathbf s_B^2,\ \mathbf x(6) = \mathbf s_C^2
\end{eqnarray}
where $\mathbf s_i^1$ and $\mathbf s_i^2$ ($i=A,B,C$) are $3 \times 1$ symbol vectors (referred to as the order-1 messages) intended to user-$i$. {As in the 2-user case, we do not introduce channel dependent precoding in the first phase in order to preserve symmetry across the users. Instead, feedback based precoding is introduced in the second phase.}

Phase-2 involves 3 time slots, in each of which two order-2 messages (defined as a combination of two order-1 messages) are sent from the first two transmit antennas:
\begin{eqnarray}
  \mathbf x(7) = \begin{bmatrix} u_{AB}^1 \\ u_{AB}^2 \\ 0 \end{bmatrix}, \ \mathbf x(8) = \begin{bmatrix} u_{AC}^1 \\ u_{AC}^2 \\ 0 \end{bmatrix},\ \mathbf x(9) = \begin{bmatrix} u_{BC}^1 \\ u_{BC}^1 \\ 0 \end{bmatrix}
\end{eqnarray}
where the order-2 messages are constructed by
\begin{eqnarray} \label{eq:msg-realization-3-user}
      u_{AB}^1 = \mathbf w_{12}^{1~T} \mathbf s_A^1 + \mathbf w_{21}^{1~T} \mathbf s_B^1 &,& u_{AB}^2 = \mathbf {w}_{12}^{2~T} \mathbf s_A^2 + \mathbf {w}_{21}^{2~T} \mathbf s_B^2 \\
      u_{AC}^1 = \mathbf w_{13}^{1~T} \mathbf s_A^1 + \mathbf w_{31}^{1~T} \mathbf s_C^1 &,& u_{AC}^2 = \mathbf {w}_{13}^{2~T} \mathbf s_A^2 + \mathbf {w}_{31}^{2~T} \mathbf s_C^2 \\
      u_{BC}^1 = \mathbf w_{23}^{1~T} \mathbf s_B^1 + \mathbf w_{32}^{1~T} \mathbf s_C^1 &,& u_{BC}^2 = \mathbf {w}_{23}^{2~T} \mathbf s_B^2 + \mathbf {w}_{32}^{2~T} \mathbf s_C^2  \label{eq:msg-realization-3-user-end}
\end{eqnarray}
where $u_{ij}^1$ and $u_{ij}^2$ $(i\ne j)$ are two realizations of the order-2 message dedicated to user-$i$ and user-$j$, and $\mathbf w_{ji}^1 \in \mathbb{C}^{3 \times 1}, \mathbf w_{ji}^2 \in \mathbb{C}^{3 \times 1}, 1 \le i,j \le 3$ can be arbitrary vector functions of $\mathbf h_i(t), i=A,B,C,~t=1,\cdots,6$. The responsibility of phase-2 is to provide independent equations with regard to $\mathbf s_i^1$ (or $\mathbf s_i^2$) by utilizing the overheard interferences in the previous phase.

Finally, in the last phase, {channel dependent precoding is not introduced as this allows to obtain decoupled optimization problems for each of the $\mathbf w_{ji}^l$ as will be made in Section III}. In this phase, two order-3 messages sent at the first transmit antenna within two consecutive time slots, i.e.,
\begin{eqnarray}
  \mathbf x(10) = \begin{bmatrix} u_{ABC}^1 \\ 0 \\ 0 \end{bmatrix}, \ \mathbf x(11) = \begin{bmatrix} u_{ABC}^2 \\ 0 \\ 0 \end{bmatrix}
\end{eqnarray}
where $u_{ABC}^l$ ($l=1,2$) is the order-3 messages which are identical to the original MAT algorithm
{
\begin{eqnarray} \nonumber
        u_{ABC}^l &=& a_{1}^l (h_{C1}(7) {u_{AB}^1}+h_{C2}(7) {u_{AB}^2}) + a_{2}^l (h_{B1}(8) {u_{AC}^1}+h_{B2}(8) {u_{AC}^2}) + a_{3}^l (h_{A1}(9) {u_{BC}^1}+h_{A2}(9) {u_{BC}^2})
\end{eqnarray}}
where $\{a_{j}^l\}$ $(j=1,2,3)$ are {chosen in a way similar to the original MAT}, i.e., arbitrary yet linearly independent sets of coefficients and known by both transmitter and receivers.

%
%

Without loss of generality, we treat user A as the target user, and the compact received signal model in matrix format over the 11 time slots can be given by
\begin{eqnarray}
\bar{\mathbf y}_A = \sqrt{\frac{P}{3}} \sum_{l=1}^2 \bar{\mathbf H}_{A1}^l \mathbf s_A^l + \sqrt{\frac{P}{3}} \sum_{l=1}^2 \bar{\mathbf H}_{A2}^l \mathbf s_B^l + \sqrt{\frac{P}{3}} \sum_{l=1}^2 \bar{\mathbf H}_{A3}^l \mathbf s_C^l + \mathbf n_A
\end{eqnarray}
where the equivalent channel matrix can be formulated as
\begin{eqnarray}
\bar{\mathbf H}_{A1}^l = \begin{bmatrix} \tilde{\mathbf H}_{A1}^l \\ \mathbf D_{A}^l{(2)} \mathbf W_1^l{(2)} \\  \mathbf D_{A}^l{(3)} \mathbf W_1^l{(3)} \end{bmatrix},
\bar{\mathbf H}_{A2}^l = \begin{bmatrix} \tilde{\mathbf H}_{A2}^l \\ \mathbf D_{A}^l{(2)} \mathbf W_2^l{(2)} \\ \mathbf D_{A}^l{(3)} \mathbf W_2^l{(3)} \end{bmatrix},
\bar{\mathbf H}_{A3}^l = \begin{bmatrix} \tilde{\mathbf H}_{A3}^l \\ \mathbf D_{A}^l{(2)}  \mathbf W_3^l{(2)} \\  \mathbf D_{A}^l{(3)}  \mathbf W_3^l{(3)} \end{bmatrix}  \in \mathbb{C}^{11 \times 3}
\end{eqnarray}
where
\begin{align}
\tilde{\mathbf H}_{Aj}^l = \begin{bmatrix} \mathbf 0_{m_{1}^l \times 3} \\  \mathbf h_A(m_{1}^l+1) \\ \mathbf 0_{n_{1}^l\times 3} \end{bmatrix} \in \mathbb{C}^{6 \times 3}
\end{align}
where $m_{1}^l=(3(l-1)+j-1)$, $n_{1}^l=6-3(l-1)-j$ and $\mathbf D_{A}^l{(2)}=\text{diag} \{h_{Al}(7),h_{Al}(8), h_{Al}(9)\}$, $\mathbf D_{A}^l{(3)}=\text{diag}\{h_{A1}(10), h_{A1}(11)\}$, and
\begin{align}
  \mathbf W^l{(2)}=  \begin{bmatrix} \underbrace{\Bmatrix{\mathbf w_{12}^{l~T} \\ \mathbf w_{13}^{l~T} \\  \mathbf 0_{1\times3}  }}_{\mathbf W_1^l{(2)}} & \underbrace{\Bmatrix{\mathbf w_{21}^{l~T} \\\mathbf 0_{1\times3} \\ \mathbf w_{23}^{l~T} }}_{\mathbf W_2^l{(2)}} & \underbrace{\Bmatrix { \mathbf 0_{1\times3} \\\mathbf w_{31}^{l~T} \\ \mathbf w_{32}^{l~T} }}_{\mathbf W_3^l{(2)}} \end{bmatrix}\in \mathbb{C}^{3 \times 9}
\end{align}
is the global precoding matrix (which is referred to hereafter as the order-2 message generation matrix) and $\mathbf W_j^l{(2)}$ is corresponding to user-$j$.

Given the order-2 message generation matrix $\mathbf W_j^l{(2)} \in \mathbb{C}^{3 \times 3}$, the precoding matrix for the third phase (referred to as order-3 message generation matrix) can be recursively obtained by
\begin{eqnarray}
\mathbf W_j^l{(3)}=\mathbf C^l(2) {\bf \Lambda}^l(2) \mathbf W_j^l{(2)}\in \mathbb{C}^{2 \times 3}, \quad j=1,2,3
\end{eqnarray}
where $\mathbf {\bf \Lambda}^l(2) = \text{diag} \{ h_{Cl}(7), h_{Bl}(8), h_{Al}(9) \}$ is set identically to MAT for simplicity, and
\begin{align}
\mathbf C^l(2) = \begin{pmatrix} a_{1}^1 & a_{2}^1 & a_{3}^1 \\ a_{1}^2 & a_{2}^2 & a_{3}^2 \end{pmatrix}
\end{align}
is a constant matrix known by both transmitter and receivers.

\subsubsection{A Particular Case (MAT Algorithm)}
The original MAT algorithm can be deduced from the proposed method by selecting
\begin{eqnarray} \label{eq:w-MAT}
\mathbf W^1{(2)} = \begin{bmatrix} \mathbf h_B^T(1)  & \mathbf h_A^T(2)  & \mathbf 0_{1\times3} \\  \mathbf h_C^T(1) & \mathbf 0_{1\times3}  & \mathbf h_A^T(3)  \\ \mathbf 0_{1\times3}  & \mathbf h_C^T(2) &  \mathbf h_B^T(3)  \end{bmatrix}
\end{eqnarray}
where $\mathbf W^2{(2)}$ can be obtained in an analogous way.

Similarly to the 2-user case, interferences carrying unintended symbols $\mathbf s_B^l$ and $\mathbf s_C^l$ are aligned perfectly at user A, {and hence matrices $\bar{\mathbf H}_{A2}^l$ and $\bar{\mathbf H}_{A3}^l$ are rank deficient with total rank of 5},  making the useful symbol $\mathbf s_A^l$ retrievable {with the left 6-dimensional interference-free subspace}. For the proposed GMAT algorithm, we seek to balance signal orthogonality {(conditioning of $\bar{\mathbf H}_{A1}^l$)} and perfect interference alignment by a careful design of $\mathbf W^l{(2)}$.

\subsection{GMAT for the General $K$-user Case}
In $K$-user case, the maximum achievable DoF is $d=\frac{K}{\sum_{k=1}^K \frac{1}{k}}$~\cite{Maddah-Ali:10}. Let $d=\frac{K^2L}{T}$, where $T$ is an integer representing the overall required time slots and $L$ is the number of repeated transmission to guarantee $T$ to be an integer. Without loss of generality, we assume $L=(K-1)!$. The total $T$ times slots can be divided into $K$ phases. In phase-1, there consists of $LK$ time slots. As the same way to the MAT algorithm, an order-1 messages $\mathbf x(t)$ is sent in $t$-th time slot, i.e.,
\begin{eqnarray}
  \mathbf x(t) = \mathbf s_i^l, \ l=1,\cdots,L
\end{eqnarray}
satisfying $t=L(l-1)+i$, where $\mathbf s_i^l$ is the $K \times 1$ symbol vector intended to user-$i$.

From phase-2 to phase-$K$, the transmission of GMAT is similar to MAT algorithm. Each phase-$k$ ($2 \le k \le K$) requires $T_k \defeq \frac{LK}{k}$ time slots, with each time slot transmitting $k$ order-$k$ message from $k$ transmit antennas, i.e.,
\begin{eqnarray}
  \mathbf x(t) = \begin{bmatrix} u_{\mathcal{S}_k}^1 & \cdots & u_{\mathcal{S}_k}^k & 0 & \cdots & 0 \end{bmatrix}^T
\end{eqnarray}
where $u_{\mathcal{S}_k}^j$ ($1\le j \le k$) is the $j$-th message realization of the order-$k$ message, which can be generated by
\begin{eqnarray}
  \mathbf u_{\mathcal{S}_k}^l = \mathbf W^l{(k)} \mathbf s^l
\end{eqnarray}
where $\mathbf u_{\mathcal{S}_k}^l$ is the $Q_k \times 1$ vector $(Q_k \defeq \binom{K}{k})$ with each element being order-$k$ message that can be interpreted as the combination of any $k$ symbol vectors from $\{ \mathbf s_i^l\}$ ($1\le l \le L$); $\mathcal{S}_k$ is the set of dedicated users and satisfies $|\mathcal{S}_k|=k$; $\mathbf s^l = [\mathbf s_1^{l~T} \ \cdots \ \mathbf s_K^{l~T}]^T \in \mathcal{C}^{K^2 \times 1}$ is the concatenated symbol vector, and $\mathbf W^l{(k)} \in \mathcal{C}^{Q_k \times K^2}$ is the order-$k$ message generation matrix, whose definition is as follows:



\begin{Definition}[\textbf{Order-$k$ Message Generation Matrix}]
  The order-$k$ message generation matrix $\mathbf W^l{(k)}=\begin{bmatrix} \mathbf W_1^l{(k)} & \cdots & \mathbf W_{K}^l{(k)}  \end{bmatrix}$ ($2\le k \le K$) is a ${Q_k \times K^2}$ matrix which satisfies:
  \begin{enumerate}
    \item it contains $k$ nonzero and $K-k$ zero blocks in each row, where each block is $1 \times K$ row vector;
    \item the positions of nonzero blocks of any two rows are not identical; and
    \item it contains all possibilities of $k$ nonzero positions out of total $K$ positions in each row.
  \end{enumerate}
\end{Definition}
We point out that the order-$k$ message is desired by those $k$ users whose symbols are contained, and acts as interferences that will be overheard by other $K-k$ users.


Based on the above definition, the signal model of $K$-user GMAT protocol can be extended as
\begin{eqnarray}
\bar{\mathbf y}_i = \sqrt{\frac{P}{K}} \sum_{l=1}^L \bar{\mathbf H}_{ii}^l \mathbf s_i^l + \sqrt{\frac{P}{K}} \sum_{l=1}^L \sum_{j=1,j \ne i}^K \bar{\mathbf H}_{ij}^l \mathbf s_j^l + \mathbf n_i
\end{eqnarray}
where
\begin{eqnarray}
\bar{\mathbf H}_{ij}^l = \begin{bmatrix} \tilde{\mathbf H}_{ij}^l(1)  \\ \vdots \\ \tilde{\mathbf H}_{ij}^l(k)  \\ \vdots \\ \tilde{\mathbf H}_{ij}^l(K) \end{bmatrix} \in \mathbb{C}^{T \times K}  \label{eq:channel-matrix-K-user}
\end{eqnarray}
with $T=\sum_{i=1}^K T_k$, is defined as follows:
\begin{itemize}
  \item The first submatrix corresponds to the effective channel matrix in phase-1, which can be given by
  \begin{align}
    \tilde{\mathbf H}_{ij}^l (1) = \begin{bmatrix} \mathbf 0_{m_1^l \times K} \\  \mathbf h_i(t) \\ \mathbf 0_{n_1^l\times K} \end{bmatrix} \in \mathbb{C}^{T_1 \times K}
  \end{align}
where $j=1,\dots,K$, $l=1,\dots, L$, $m_{1}^l=(K(l-1)+j-1)$, $n_{1}^l=KL-K(l-1)-j$, and $t=m_{1}^l+1$;
  \item The $k$-th submatrix $(2 \le k \le K-1)$ which corresponds to phase-$k$ can be formulated as
  \begin{align}
    \tilde{\mathbf H}_{ij}^l (k) = \begin{bmatrix}\mathbf 0_{m^{l}_k \times K} \\ \mathbf D_{i}^l{(k)} \mathbf W_j^l{(k)} \\ \mathbf 0_{n^{l}_k \times K} \end{bmatrix}\in \mathbb{C}^{T_k \times K}
  \end{align}
  where $m_{k}^{l} = \left(\lceil \frac{l \cdot l_k}{L}\rceil-1\right) Q_k$, $n_{k}^{l} = T_k-\lceil \frac{l\cdot l_k}{L}\rceil Q_k$ with $l_k=\frac{T_k}{Q_k}$, and $\mathbf D_{i}^l{(k)}=\text{diag}\{h_{is}(t)\} \in \mathbb{C}^{Q_k \times Q_k}$ corresponds to the present channel over whom the order-$k$ message is sent in phase-$k$ with $s=((l \cdot l_k) \mod L) \mod k$ and $t$ being the index of time slots. In general, $\mathbf W_j^l{(k)}$ ($k \ge 2$) is the order-$k$ message generation matrix specified to user-$j$, which is recursively defined according to
\begin{eqnarray}
\mathbf W_j^l{(k+1)} = \mathbf C^l(k) {\bf \Lambda}^l(k)  \mathbf W_j^l{(k)} \label{eq:message-gen}
\end{eqnarray}
where $\mathbf C^l(k)\in \mathbb{C}^{Q_{k+1} \times Q_k}$ is a constant matrix known by transmitter and all users, satisfying: (1) each row contains $k+1$ nonzero elements, and (2) the positions of nonzero elements of any two rows are different one another; and ${\bf \Lambda}^l(k) \in \mathbb{C}^{Q_k \times Q_k}$ is a diagonal matrix whose elements are chosen to be a function of the channel coefficients in phase-$k$, so that the interference overheard can be aligned within a limited dimensional subspace.  For simplicity, we place emphasis on $\mathbf W_j^l{(k)}$, letting ${\bf \Lambda}^l(k)$ be predetermined as the channel coefficients in phase-$k$ like the original MAT algorithm.
  \item The last submatrix is corresponding to the last phase, i.e.,
  \begin{align}
    \tilde{\mathbf H}_{ij}^l (K) = \mathbf D_{i}^l{(K)} \mathbf W_j^l{(K)} \in \mathbb{C}^{T_K \times K}
  \end{align}
  where $\mathbf W_j^l{(K)}$ is defined similarly to (\ref{eq:message-gen}), in which $\mathbf C^l(K-1) \in \mathbb{C}^{T_{K} \times Q_{K-1}} $ is a full rank constant matrix without zero elements, and $\mathbf D_{i}^l{(K)} =\text{diag}\{h_{i1}(t)\} \in \mathbb{C}^{T_{K} \times T_{K}}$ contains channel coefficients during phase-$K$.
\end{itemize}


For further illustration, we take the 4-user case for example to show its order-2 message generation matrix, i.e.,
\begin{eqnarray}
\mathbf W^l{(2)}=\begin{bmatrix}
    \mathbf w_{12}^{l~T} & \mathbf w_{21}^{l~T} & \mathbf 0 & \mathbf 0 \\
    \mathbf w_{13}^{l~T} & \mathbf 0 & \mathbf w_{31}^{l~T} & \mathbf 0 \\
    \mathbf w_{14}^{l~T} & \mathbf 0 & \mathbf 0 & \mathbf w_{41}^{l~T} \\
    \mathbf 0 & \mathbf w_{23}^{l~T} & \mathbf w_{32}^{l~T} & \mathbf 0 \\
    \mathbf 0 & \mathbf w_{24}^{l~T} & \mathbf 0 & \mathbf w_{42}^{l~T} \\
    \mathbf 0 & \mathbf 0 & \mathbf w_{34}^{l~T} & \mathbf w_{43}^{l~T}
  \end{bmatrix}
\end{eqnarray}
where $\mathbf w_{ji}^l \in \mathbb{C}^{K \times 1}$ is the beamforming vector aiming at the compromise between user-$i$ and user-$j$. {This formulation collapses to (\ref{eq:msg-realization-3-user})-(\ref{eq:msg-realization-3-user-end}) for the 3-user case and to (\ref{eq:msg-realization-2-user}) for the 2-user case.}

\subsubsection{A particular Case (MAT Algorithm)}
Particularly for the 4-user case, the original MAT algorithm is a specialized GMAT algorithm by setting order-2 message generation matrix as
\begin{eqnarray}
  \mathbf W^1{(2)}=\begin{bmatrix}
    \mathbf h_B^T(1) & \mathbf h_A^T(2) & \mathbf 0 & \mathbf 0 \\
    \mathbf h_C^T(1) & \mathbf 0 & \mathbf h_A^T(3) & \mathbf 0 \\
    \mathbf h_D^T(1) & \mathbf 0 & \mathbf 0 & \mathbf h_A^T(4) \\
    \mathbf 0 & \mathbf h_C^T(2) & \mathbf h_B^T(3) & \mathbf 0 \\
    \mathbf 0 & \mathbf h_D^T(2) & \mathbf 0 & \mathbf h_B^T(4) \\
    \mathbf 0 & \mathbf 0 & \mathbf h_D^T(3) & \mathbf h_C^T(4)
  \end{bmatrix}
\end{eqnarray}
for $l=1$ and similarly for other $l$.
For example, for user A, the interference channels $\bar{\mathbf H}_{Aj}^l$ ($j \ne 1$) are perfectly aligned, leaving $K=4$ interference free dimensions for desired signal, and therefore making the intended symbols retrievable at user A. Similarly for other users, all symbols can be recovered. Hence, $96$ symbols are delivered within $50$ time slots, providing the sum DoF of $\frac{48}{25}$.

It is worth noting that the higher level messages can be delivered by the combination of lower lever messages. For example, from phase $k$ to $K$, the message delivered to the receivers aiming at completely decoding the order-$k$ message. To avoid too many parameters being optimized which requires huge complexity, we will focus merely on the design of the order-2 message generation matrices $\{\mathbf W_j^l{(2)}\}$.

\section{GMAT Optimization Design}
The computation of $\{\mathbf W_j^l{(2)}\}$ can use several options. Two of them are briefly described in the following sections. The first is based on the optimization of a virtual MMSE metric, yielding an iterative solution, while the second one considers the maximization of an approximation of the mutual information, yielding suboptimal yet closed-form solutions. Note that none of these approaches have anything in common with finite SNR interference alignment methods with non-delayed CSIT, such as, e.g.,~\cite{Gomadam:11,Peters:09,Negro:10}, since the nature of our problem is conditioned by the delayed CSIT scenario.

\subsection{Virtual MMSE Metric}
In the following, we describe an approach based on a virtual MMSE metric (referred to later as ``GMAT-MMSE'') for the 2-user case, and subsequently generalize it to the $K$-user case.
\subsubsection{Special $K=2$ Case}
Since the transmitter does not know ${\bf h}_i(3)$ at slot-$3$, the optimization of the precoder in (\ref{eq:channel-vector-GMAT-A}) and (\ref{eq:channel-vector-GMAT-B}) cannot involve such information. Fortunately, we point out that the trade-off between interference alignment and signal matrix orthogonalization presented above can be formulated in a way that is fully independent of ${\bf h}_i(3)$. To do so, we introduce the virtual received signal ${\mathbf y}_i$ given below, where ${\bf h}_i(3)$ is ignored (deterministic fading is assumed over the third time slot):
\begin{eqnarray}
  {\mathbf y}_i = \sqrt{\frac{P}{2}}   {\mathbf H}_{i1} \mathbf s_A + \sqrt{\frac{P}{2}}  {\mathbf H}_{i2} \mathbf s_B +   \mathbf n_i , i=A,B
\end{eqnarray}
where the virtual channel matrices are now modified from (\ref{eq:channel-vector-GMAT-A}) and (\ref{eq:channel-vector-GMAT-B}) by simply setting ${h}_{i1}(3)=1$:
\begin{eqnarray} \label{eq:virtual-channel}
 \mathbf H_{i1}=\begin{bmatrix} \mathbf h_i^T(1) \\ \mathbf 0 \\ \mathbf w_1^T \end{bmatrix}, \  \mathbf H_{i2}=\begin{bmatrix} \mathbf 0 \\ \mathbf h_i^T(2) \\ \mathbf w_2^T
 \end{bmatrix}, \  i=A,B.
\end{eqnarray}
Given $\mathbf w_1$ and $\mathbf w_2$, the optimum RX MMSE filters at user-$i$ over this channel are given by
\begin{eqnarray} \label{eq:MMSE-A}
 \mathbf V_i  &=&\sqrt{\rho} \left( \rho \mathbf H_{i1}\mathbf H_{i1}^H+\rho\mathbf H_{i2}\mathbf H_{i2}^H+ \mathbf I\right)^{-1} \mathbf H_{i1}
\end{eqnarray}
where $\rho=\frac{P}{K}$ (here $K=2$), and the corresponding optimal MSEs are
\begin{eqnarray} \label{eq:sum-MSE-A}
J_i(\mathbf w_1,\mathbf w_2)&=&\text{Tr}\left(\mathbf I - \rho \mathbf H_{i1}^H (\rho\mathbf H_{i1}\mathbf H_{i1}^H+\rho\mathbf H_{i2}\mathbf H_{i2}^H+  \mathbf I)^{-1}\mathbf H_{i1} \right)
\end{eqnarray}

Hence, the optimal $\mathbf w_1$,$\mathbf w_2$ can be obtained from the following optimization problem, i.e.,
\begin{eqnarray} \label{eq:MMSE-Obj-2-user}
 \min_{\mathbf w_1,\mathbf w_2: \|\mathbf w_1\|^2+\|\mathbf w_2\|^2 \le 2} && J= J_A(\mathbf w_1,\mathbf w_2) + J_B(\mathbf w_1,\mathbf w_2)
\end{eqnarray}
{In practice, the gradient based approaches can be used to perform optimization although the convexity of the problem is not guaranteed.}

\subsubsection{General $K$-user Case}
In phase-$k$, the transmitter does not know ${\bf h}_i(t)$ at slot-$t$, where $t=\sum_{l=1}^{k-1}T_{l}+1, \cdots, \sum_{l=1}^{k}T_{l}$. Similarly to the 2-user case, the virtual received signal can be generalized as
\begin{eqnarray}
{\mathbf y}_i = \sqrt{\frac{P}{K}} \sum_{l=1}^L {\mathbf H}_{ii}^l \mathbf s_i^l + \sqrt{\frac{P}{K}} \sum_{l=1}^L \sum_{j=1,j \ne i}^K {\mathbf H}_{ij}^l \mathbf s_j^l + \mathbf n_i,~i=1,\cdots,K
\end{eqnarray}
where
\begin{eqnarray}
{\mathbf H_{ij}^l}=
\begin{bmatrix} \tilde{\mathbf H}_{ij}^{l~T}  & \cdots & \mathbf 0_{K \times m_k^{l}} & \mathbf W_j^{l~T}{(k)} & \mathbf 0_{K \times n_k^{l}} & \cdots & \mathbf W_j^{l~T}{(K)} \end{bmatrix}^T
\end{eqnarray}
whose elements are defined in Section II.

Similarly, given $\mathbf W_j^l{(2)}$, the optimum MMSE filters for $\mathbf s_i^l$ at user-$i$ becomes
\begin{eqnarray} \label{eq:MMSE-A}
 \mathbf V_i^l  &=&\sqrt{\rho} \left( \rho \sum_{l=1}^L \sum_{j=1}^K \mathbf H_{ij}^l\mathbf H_{ij}^{l~H} + \mathbf I\right)^{-1} \mathbf H_{ii}^l
\end{eqnarray}
where $\rho=\frac{P}{K}$ is the normalized transmit power, and the corresponding optimal MSEs are
\begin{eqnarray}
J_i^l(\mathbf W_j^l{(2)}, j=1,\cdots,K)=\text{Tr}\left(\mathbf I - \rho \mathbf H_{ii}^{l~H} \left(\rho \sum_{l=1}^L \sum_{j=1}^K \mathbf H_{ij}^l\mathbf H_{ij}^{l~H} + \mathbf I \right)^{-1}\mathbf H_{ii}^l \right)
\end{eqnarray}

The optimal solutions of $\{\mathbf W_j^l{(2)},j=1,\cdots,K\}$ in the sense of virtual MMSE at receiver side are now given by:
\begin{eqnarray}
 \min_{\mathbf W_j^l{(2)}, j=1,\cdots,K} && J= \sum_{l=1}^L \sum_{i=1}^K J_i^l(\mathbf W_j^l{(2)}) \\
 s.t.&&\sum_{l=1}^L \sum_{j=1}^K \| \mathbf W_j^l{(2)} \|^2_F \le KT_2.
\end{eqnarray}
As the above optimization does not lend itself easily to a closed-form solution, we propose an iterative procedure, based on the gradient descent of the cost function $J$, where ${\mathbf W}_j^l{(2)}$ is iterative updated according to
\begin{eqnarray}
\hat{\mathbf W}_j^l{(2)}[n+1]= \hat{\mathbf W}_j^l{(2)}[n] - \beta \frac{\partial (J)}{\partial \mathbf W_j^l{(2)}}
\end{eqnarray}
where $n$ is the iteration index and $\beta$ is a small step size. The partial derivation is given in the Appendix. Nevertheless, to circumvent non-convexity issues, we explore an alternative optimization method below.

\subsection{Mutual Information Metric}
Here, we propose an approach based on maximizing an approximation of the mutual information, yielding a convenient closed-form solution for $\{\mathbf W_j^l{(2)}\}$. In the following, we will start with the 2-user case to gain insight, and then generalize it to the $K$-user case.
\subsubsection{Special 2-user Case}
Recall that
\begin{eqnarray}
  \mathbf y_A&=& \sqrt{\rho} {\bar{\mathbf H}_{A1}} \mathbf s_A + \sqrt{\rho} {\bar{\mathbf H}_{A2}} \mathbf s_B + \mathbf n_A
\end{eqnarray}
where $\rho=\frac{P}{K}$ (here $K=2$), $\mathbf w_1$ and $\mathbf w_2$ are functions of $\mathbf h_i(j), i=A,B, j=1,2$ and satisfy power constraint $\|\mathbf w_1\|^2 + \|\mathbf w_2\|^2 \le 2$. Consequently, the exact mutual information of user A can be calculated by
{
\begin{eqnarray}
  I(\mathbf s_A ; \mathbf y_A) &=& \log \det \left( \mathbf I + \left( \mathbf I + \rho \bar{\mathbf H}_{A2} \bar{\mathbf H}_{A2}^H \right)^{-1} \rho \bar{\mathbf H}_{A1} \bar{\mathbf H}_{A1}^H \right)\\
  &=& \log \det \left( \mathbf I +  \rho \begin{bmatrix} 1 & 0 \\ 0 & \frac{1+\|\mathbf h_{A}^H(2)\|^2}{ \Delta_1(\mathbf w_2)}   \end{bmatrix} \begin{bmatrix} \|\mathbf h_{A}^H(1)\|^2 & h^*_{A1}(3)  \mathbf w_1^H \mathbf h_{A}(1)\\ h_{A1}(3) \mathbf h_{A}^H(1) \mathbf w_1 & |h_{A1}(3)|^2 \|\mathbf w_1\|^2\end{bmatrix} \right)\\
  &=& \log \left( 1+\rho\|\mathbf h_A(1)\|^2 + \frac{\Theta_1(\mathbf w_1)}{\Delta_1(\mathbf w_2)}  \right)
\end{eqnarray}
}
where the second line is easily obtained by permuting rows 2 and 3 in $\bar{\mathbf H}_{A1}$ and $\bar{\mathbf H}_{A2}$, and the third line by the characteristic polynomial equality~\cite{Aitken:56},
$  \det (\mathbf I + \rho \mathbf M) = 1+\rho~\text{\text{Tr}}(\mathbf M) + \rho^2 \det (\mathbf M)$,
where $\mathbf M$ is a $2 \times 2$ Hermitian matrix.
By analogy, the mutual information of user B can be given by
\begin{eqnarray} \label{eq:MI-equation}
  I(\mathbf s_B ; \mathbf y_B) &=& \log \left( 1+\rho\|\mathbf h_B(2)\|^2 + \frac{\Theta_2(\mathbf w_2)}{\Delta_2(\mathbf w_1)}  \right)
\end{eqnarray}
where
{
\begin{eqnarray}
 \Theta_1(\mathbf w_1) &=& (1+\rho\|\mathbf h_A(2)\|^2) \rho |h_{A1}(3)|^2 (\|\mathbf w_1\|^2 + \rho \|\mathbf w_1\|^2 \|\mathbf h_A(1)\|^2 - \rho \mathbf w_1^H \mathbf h_A(1) \mathbf h_A(1)^H \mathbf w_1)\\
  \Delta_1(\mathbf w_2)&=&(1+\rho\|\mathbf h_A(2)\|^2)(1+\rho |h_{A1}(3)|^2 \|\mathbf w_2\|^2) - \rho^2 |h_{A1}(3)|^2 \mathbf w_2^H \mathbf h_A(2) \mathbf h_A(2)^H \mathbf w_2\\
  \Theta_2(\mathbf w_2) &=& (1+\rho\|\mathbf h_B(1)\|^2) \rho |h_{B1}(3)|^2 (\|\mathbf w_2\|^2 + \rho \|\mathbf w_2\|^2 \|\mathbf h_B(2)\|^2 - \rho \mathbf w_2^H \mathbf h_B(2) \mathbf h_B(2)^H \mathbf w_2)\\
  \Delta_2(\mathbf w_1)&=&(1+\rho\|\mathbf h_B(1)\|^2)(1+\rho |h_{B1}(3)|^2 \|\mathbf w_1\|^2) - \rho^2 |h_{B1}(3)|^2 \mathbf w_1^H \mathbf h_B(1) \mathbf h_B(1)^H \mathbf w_1
\end{eqnarray}
}
By imposing a symmetric constraint for power allocation between $\mathbf w_1$ and $\mathbf w_2$, e.g., $\|\mathbf w_1\|^2=\|\mathbf w_2\|^2=1$ for simplicity, the sum mutual information can be deduced to
\begin{eqnarray} \label{eq:sum-MI-equation}
  I(\mathbf s_A ; \mathbf y_A) + I(\mathbf s_B ; \mathbf y_B) &=&  \log \left( 1+  \frac{\mathbf w_1^H \mathbf R_1 \mathbf w_1}{\mathbf w_2^H \mathbf R_2 \mathbf w_2} \right) + \log \left( 1+  \frac{\mathbf w_2^H \mathbf Q_2 \mathbf w_2}{\mathbf w_1^H \mathbf Q_1 \mathbf w_1} \right) + \log C
\end{eqnarray}
where
{
\begin{eqnarray}
  \mathbf R_1 &=& (1+\rho\|\mathbf h_A(2)\|^2)\left(\mathbf I + \rho \mathbf h_A^{\bot}(1) \mathbf h_A^{\bot H}(1)\right) \\
  \mathbf R_2 &=& (1+\rho\|\mathbf h_A(1)\|^2) \left( \gamma_1 \mathbf I + \rho \mathbf h_A^{\bot}(2) \mathbf h_A^{\bot H}(2) \right)\\
   \mathbf Q_1 &=& (1+\rho\|\mathbf h_B(2)\|^2) \left( \gamma_2 \mathbf I + \rho \mathbf h_B^{\bot}(1) \mathbf h_B^{\bot H}(1) \right) \\
  \mathbf Q_2 &=& (1+\rho\|\mathbf h_B(1)\|^2) \left(\mathbf I + \rho \mathbf h_B^{\bot}(2) \mathbf h_B^{\bot H}(2)\right)
\end{eqnarray}
}
where
\begin{eqnarray}
  &&\gamma_1=\frac{1+\rho\|\mathbf h_A(2)\|^2}{\rho |h_{A1}(3)|^2} + \|\mathbf w_2\|^2, \gamma_2= \frac{1+\rho\|\mathbf h_B(1)\|^2}{\rho |h_{B1}(3)|^2} + \|\mathbf w_1\|^2, \\
  &&C = (1+\rho\|\mathbf h_A(1)\|^2)(1+\rho\|\mathbf h_B(2)\|^2)
\end{eqnarray}
and $\mathbf h_i^{\bot}(j) \in \mathbb{C}^{2 \times 1}$ is the orthogonal channel of $\mathbf h_i(j)$ ($i=A,B, j=1,2$) satisfying
\begin{eqnarray}
  \mathbf h_i(j) \mathbf h_i^H(j) + \mathbf h_i^{\bot}(j) \mathbf h_i^{\bot H}(j) = \|\mathbf h_i(j)\|^2 \mathbf I.
\end{eqnarray}

In the high SNR region, we get a useful approximation of the sum of mutual informations, i.e.,
\begin{eqnarray}
I(\mathbf s_A ; \mathbf y_A) + I(\mathbf s_B ; \mathbf y_B) \approx  \log \left( \frac{\mathbf w_1^H \mathbf R_1 \mathbf w_1}{\mathbf w_2^H \mathbf R_2 \mathbf w_2}  \frac{\mathbf w_2^H \mathbf Q_2 \mathbf w_2}{\mathbf w_1^H \mathbf Q_1 \mathbf w_1} \right) + \log C
\end{eqnarray}
which can be optimized by separately maximizing the two Rayleigh Quotients, i.e.,
\begin{eqnarray} \label{eq:MI-1-obj}
  \max_{\|\mathbf w_1\|^2=1} \  \frac{\mathbf w_1^H \mathbf R_1 \mathbf w_1}{\mathbf w_1^H \mathbf Q_1 \mathbf w_1} &=& \max_{\|\mathbf w_1\|^2=1} \ \frac{\mathbf w_1^H \left(\mathbf I + \rho \mathbf h_A^{\bot}(1) \mathbf h_A^{\bot H}(1)\right) \mathbf w_1}{\mathbf w_1^H \left( \gamma_2 \mathbf I + \rho \mathbf h_B^{\bot}(1) \mathbf h_B^{\bot H}(1) \right) \mathbf w_1}\\ \label{eq:MI-1-obj0}
  \max_{\|\mathbf w_2\|^2 = 1} \  \frac{\mathbf w_2^H \mathbf Q_2 \mathbf w_2}{\mathbf w_2^H \mathbf R_2 \mathbf w_2} &=& \max_{\|\mathbf w_2\|^2 = 1} \  \frac{\mathbf w_2^H \left(\mathbf I + \rho \mathbf h_B^{\bot}(2) \mathbf h_B^{\bot H}(2)\right) \mathbf w_2}{\mathbf w_2^H \left( \gamma_1 \mathbf I + \rho \mathbf h_A^{\bot}(2) \mathbf h_A^{\bot H}(2) \right) \mathbf w_2}
\end{eqnarray}
{Hence, we can obtain the optimal solutions $\mathbf w_1^{opt}$ and $\mathbf w_2^{opt}$, which are given by} the dominant generalized eigenvectors of the pairs $(\mathbf R_1,\mathbf Q_1)$ and $(\mathbf Q_2,\mathbf R_2)$, respectively.

{Interestingly, the above objective function can be interpreted as dual SINR in a 2-user interference channel.} Define
\begin{eqnarray} \label{eq:Def-DSINR}
\text{DSINR}_i = \frac{\mathbf w_i^H \left(\mathbf I + \rho \mathbf h_i^{\bot}(i) \mathbf h_i^{\bot H}(i)\right) \mathbf w_i}{\mathbf w_i^H \left( \gamma_{\bar i} \mathbf I + \rho \mathbf h_{\bar i}^{\bot}(i) \mathbf h_{\bar i}^{\bot H}(i) \right) \mathbf w_i}
\end{eqnarray}
which is referred to as {a} \emph{regularized} SINR in a dual 2-user interference channel with a desired channel $\mathbf h_i^{\bot}$ and interference channel $\mathbf h_{\bar i}^{\bot}$, where $i \ne \bar i$, and $\mathbf w_i$ is interpreted as a receive filter. Thus, the optimization problem in eq-(\ref{eq:MI-1-obj}) can be equivalently done by maximizing the regularized SINR in the dual MISO interference channels. Note that the regularization lies in not only the interference channels but also the desired channels. This solution is referred to later as ``GMAT-DSINR''.

\subsubsection{General $K$-user Case}
Recall that the definition of DSINR in eq-(\ref{eq:Def-DSINR}) for the 2-user case, where $\mathbf w_i$ is determined by the orthogonal channels of itself and also its peer. According to the structure of $\mathbf W^l{(2)}$ for the $K$-user case, we can follow this approach and design each nonzero submatrices $\mathbf w_{ji}^l$ distributively. For each $\mathbf w_{ji}^l$, the dual interference channel can be constructed by the orthogonal channels between itself $\mathbf h_j^{\bot}$ and its peer $\mathbf h_i^{\bot}$. Thus, the regularized dual SINR can be formulated as (e.g., $l=1$)
\begin{eqnarray} \label{eq:DSINR-general}
 \text{DSINR}_{ji}^l &=& \frac{\mathbf w_{ji}^{l~H} \left(\mathbf I + \rho \sum_{k \neq i} \mathbf h_k^{\bot}(j) \mathbf h_k^{\bot H}(j)\right) \mathbf w_{ji}^l}{\mathbf w_{ji}^{l~H} \left( \gamma_{ji} \mathbf I + \rho \mathbf h_{i}^{\bot}(j) \mathbf h_{i}^{\bot H}(j) \right) \mathbf w_{ji}^l}, ~~~ j \ne i
\end{eqnarray}
where $\mathbf w_{ji}^l \in \mathbb{C}^{K \times 1}$ is the $i$-th (when $i<j$) or $(i-1)$-th (when $i>j$) nonzero block of $\mathbf W_j^l{(2)}$, $\mathbf h_i^{\bot}(j) \in \mathbb{C}^{K \times K}$ is one representation of the null space of $\mathbf h_i(j)$ with the same norm\footnote{We abuse here the vector notation to represent the corresponding orthogonal channel matrix for the sake of consistence.}, and
\begin{eqnarray}
  \gamma_{ji} =  \|\mathbf w_{ji}^l\|^2 + \|\mathbf h_i(j)\|^2 + 1/\rho
\end{eqnarray}
Accordingly, the optimal $\mathbf w_{ji}^l$ can be obtained by distributively optimizing
\begin{eqnarray} \label{eq:K-user-DSINR-O}
  \max_{\mathbf w_{ji}^l}&& \{\text{DSINR}_{ji}^l,~j \ne i\}\\ \label{eq:K-user-DSINR-C}
  s.t.&&\sum_{l=1}^L \sum_{j=1}^K \| \mathbf W_j^l{(2)} \|^2_F \le KT_2.
\end{eqnarray}
where the corresponding solution can be simply obtained by generalized eigenvalue decomposition. By maximizing the dual SINR, $\mathbf w_{ji}^l$ is preferred to keep aligned along with $\mathbf h_j(j)$ while to be as orthogonal to $\mathbf h_k(j)$ as possible. Consequently, the optimal solution of $\mathbf w_{ji}^l$ balances signal orthogonality with interference alignment between user-$j$'s and other users' dual orthogonal channels at $j$-th time slot.

\section{Discussion}
\subsection{Multiplexing Gain of GMAT}
In the following, we show the GMAT algorithm possesses the same multiplexing gain as original MAT. We consider the 2-user case for example. According to equations from (\ref{eq:sum-MI-equation}) to (\ref{eq:MI-1-obj0}), we have
\begin{eqnarray}
&\lim_{\rho \to \infty} \frac{\mathbb{E}{\log \left(\max_{\|\mathbf w_1\|^2=1} \frac{\mathbf w_1^H \mathbf R_1 \mathbf w_1}{\mathbf w_1^H \mathbf Q_1 \mathbf w_1} \right)}} { \log \rho} = \lim_{\rho \to \infty} \frac{\mathbb{E}{\left.\log \left(\frac{\mathbf w_1^H \mathbf R_1 \mathbf w_1}{\mathbf w_1^H \mathbf Q_1 \mathbf w_1} \right) \right|_{\mathbf w_1 = \frac{\mathbf h_B(1)}{\|\mathbf h_B(1)\|}}}}  { \log \rho} = 1\\
&\lim_{\rho \to \infty} \frac{\mathbb{E}{\log \left(\max_{\|\mathbf w_2\|^2=1}  \frac{\mathbf w_2^H \mathbf Q_2 \mathbf w_2}{\mathbf w_2^H \mathbf R_2 \mathbf w_2} \right)}}{\log \rho} = \lim_{\rho \to \infty} \frac{\mathbb{E}{\left.\log \left(\frac{\mathbf w_2^H \mathbf Q_2 \mathbf w_2}{\mathbf w_2^H \mathbf R_2 \mathbf w_2} \right) \right|_{\mathbf w_2 = \frac{\mathbf h_A(2)}{\|\mathbf h_A(2)\|}}}}  { \log \rho} = 1
\end{eqnarray}
Thus, together with the fact that $\lim_{\rho \to \infty} \frac{\mathbb{E} \log C }{ \log \rho} =2$, the multiplexing gain can be achieved with
\begin{align}
  \text{MG}_{\text{GMAT}} &= \lim_{\rho \to \infty} \frac{\mathbb{E} \max_{\|\mathbf w_1\|^2=1,\|\mathbf w_2\|^2=1}(I(\mathbf s_A ; \mathbf y_A) + I(\mathbf s_B ; \mathbf y_B))}{3 \log \rho} = \frac{4}{3}
\end{align}
which is identical to the original MAT algorithm. Intuitively, at high SNR, the signal orthogonality becomes no relevance, thus our solution naturally seeks perfect interference alignment as in MAT.

%

\subsection{Single-beam MIMO Interference Channel Interpretation}
To understand more clearly the roles of desired signal orthogonality and interference alignment, we transform the mutual information equality (\ref{eq:sum-MI-equation}) into another form, and further interpret their relationship from the point of view of a two-user single-beam MIMO interference channel. {The strong benefit of this interpretation is that the problem of computing the precoders lends itself to classical precoding techniques in the MIMO interference channel.} Based on eq-(\ref{eq:sum-MI-equation}), the sum mutual information equation can be further transformed to
\begin{eqnarray} \label{eq:MI-IFC}
  &&I(\mathbf s_A ; \mathbf y_A) + I(\mathbf s_B ; \mathbf y_B)\\ \label{eq:IFC-1}
  &=& \log \left( 1+ \frac{\alpha_1 \rho \mathbf w_1^H \mathbf h_A(1) \mathbf h_A^H(1) \mathbf w_1 + \alpha_2 \rho \mathbf w_1^H \mathbf h_A^{\bot}(1) \mathbf h_A^{\bot H}(1) \mathbf w_1}{\sigma_1^2+ \beta_3 \rho \mathbf w_2^H \mathbf h_A(2) \mathbf h_A^{H}(2) \mathbf w_2 + \beta_4 \rho \mathbf w_2^H \mathbf h_A^{\bot}(2) \mathbf h_A^{\bot H}(2) \mathbf w_2} \right) \\ \label{eq:IFC-2}
  &+& \log \left( 1+  \frac{\beta_1 \rho \mathbf w_2^H \mathbf h_B(2) \mathbf h_B^H(2) \mathbf w_2+\beta_2 \rho \mathbf w_2^H \mathbf h_B^{\bot}(2) \mathbf h_B^{\bot H}(2) \mathbf w_2}{\sigma_2^2+\alpha_3 \rho \mathbf w_1^H \mathbf h_B(1) \mathbf h_B^{H}(1) \mathbf w_1+\alpha_4 \rho \mathbf w_1^H \mathbf h_B^{\bot}(1) \mathbf h_B^{\bot H}(1) \mathbf w_1} \right)
  + \log C
\end{eqnarray}
where
\begin{eqnarray}
  &&\alpha_1=\frac{\alpha_2}{1+\rho\|\mathbf h_A(1)\|^2}, \alpha_2=\frac{1+\rho\|\mathbf h_A(2)\|^2}{\rho\|\mathbf h_A(1)\|^2}, \alpha_3=\frac{1}{\rho |h_{B1}(3)|^2 \|\mathbf w_1\|^2}, \alpha_4=\alpha_3+1,\\
  &&\beta_1=\frac{\beta_2}{1+\rho\|\mathbf h_B(2)\|^2}, \beta_2=\frac{1+\rho\|\mathbf h_B(1)\|^2}{\rho\|\mathbf h_B(2)\|^2}, \beta_3=\frac{1}{\rho |h_{A1}(3)|^2 \|\mathbf w_2\|^2}, \beta_4=\beta_3+1, \\
  &&\sigma_1^2=\frac{1}{\rho |h_{A1}(3)|^2} + \|\mathbf w_2\|^2, \sigma_2^2= \frac{1}{\rho |h_{B1}(3)|^2} + \|\mathbf w_1\|^2.
\end{eqnarray}
According to eq-(\ref{eq:IFC-1}) and eq-(\ref{eq:IFC-2}), the sum mutual information can be treated as that of 2-user MIMO interference channels with 2 antennas at each transmitter and receiver, as shown in Fig.~1. Note that $\mathbf w_1$ and $\mathbf w_2$ act as the transmit beamformers, where the single beam is transmitted from each transmitter.

Accordingly, the received signal at two receivers can be equivalently expressed as
\begin{eqnarray}
  \mathbf y_1 &=& \sqrt{\rho} \mathbf H_1 \mathbf w_1 s_1 + \sqrt{\rho} \mathbf H_2 \mathbf w_2 s_2 + \mathbf n_1\\
  \mathbf y_2 &=& \sqrt{\rho} \mathbf G_2 \mathbf w_2 s_2 + \sqrt{\rho} \mathbf G_1 \mathbf w_1 s_1 + \mathbf n_2
\end{eqnarray}
where
\begin{eqnarray}
  \mathbf H_1 = \begin{bmatrix} \sqrt{\alpha_1} \mathbf h_A^H(1) \\ \sqrt{\alpha_2} \mathbf h_A^{\bot H}(1)\end{bmatrix}, \mathbf H_2 =  \begin{bmatrix}\sqrt{\beta_3} \mathbf h_A^H(2) \\ \sqrt{\beta_4} \mathbf h_A^{\bot H}(2)\end{bmatrix},  \mathbf G_1 = \begin{bmatrix} \sqrt{\alpha_3} \mathbf h_B^H(1) \\ \sqrt{\alpha_4} \mathbf h_B^{\bot H}(1)\end{bmatrix},  \mathbf G_2 = \begin{bmatrix} \sqrt{\beta_1} \mathbf h_B^H(2)\\ \sqrt{\beta_2} \mathbf h_B^{\bot H}(2) \end{bmatrix}
\end{eqnarray}
and the noises are distributed with $\mathbf n_i \sim \mathcal{CN}(0, \frac{\sigma_i^2}{2} \mathbf I)$, respectively.

Consequently, the received SINR for two users can be written, respectively, as
\begin{eqnarray}
  \text{SINR}_1 &=& \frac{\rho\|\mathbf H_1 \mathbf w_1\|^2}{\sigma_1^2+\rho\|\mathbf H_2 \mathbf w_2\|^2} = \frac{\rho\mathbf w_1^H \mathbf H_1^H \mathbf H_1 \mathbf w_1}{\sigma_1^2+\rho\mathbf w_2^H \mathbf H_2^H \mathbf H_2 \mathbf w_2 }\\
  \text{SINR}_2 &=& \frac{\rho\|\mathbf G_2 \mathbf w_2\|^2}{\sigma_2^2+\rho\|\mathbf G_1 \mathbf w_1\|^2} = \frac{\rho\mathbf w_2^H \mathbf G_2^H \mathbf G_2 \mathbf w_2}{\sigma_2^2+\rho\mathbf w_1^H \mathbf G_1^H \mathbf G_1 \mathbf w_1 }
\end{eqnarray}
which are identical to those in eq-(\ref{eq:IFC-1}-\ref{eq:IFC-2}). Hence, existing precoder design methods in the two-user single-beam MIMO interference channels with perfect CSIT, e.g.,~\cite{Negro:10,Ho:10,Schmidt:10,Peters:11,Cao:11}, can be used here in the context of delayed CSIT precoding. Instead of going into details about those solutions, we take the classic MRT and ZF precoders here for example,
\begin{eqnarray}
  \mathbf w_1^{MRT} = \mathbf U_{\max} (\mathbf H_1^H \mathbf H_1) &,&  \mathbf w_2^{MRT} = \mathbf U_{\max} (\mathbf G_2^H \mathbf G_2)\\
  \mathbf w_1^{ZF} = \mathbf U_{\min} (\mathbf G_1^H \mathbf G_1) &,&  \mathbf w_2^{ZF} = \mathbf U_{\min} (\mathbf H_2^H \mathbf H_2)
\end{eqnarray}
where $\mathbf U_{\max}(\cdot)$ and $\mathbf U_{\min}(\cdot)$ are the generalized {eigenvectors} corresponding to the largest and smallest eigenvalues, respectively. Interestingly, for the first user, it is worth noting that $\alpha_1 < \alpha_2$ and therefore $\mathbf w_1^{MRT} \to \mathbf h_A^{\bot}(1)$, means perfect orthogonality of desired signal is preferred. On the other hand, $\alpha_3 < \alpha_4$, which denotes $\mathbf w_1^{ZF} \to \mathbf h_B(1)$, corresponds to the preference of perfect interference alignment. Our proposed GMAT-MMSE and GMAT-DSINR solutions offer a trade-off between them, yielding a better performance in finite SNR regime.

\section{Numerical Results}
The effectiveness of the proposed solutions is evaluated in terms of the sum rate per time slot in bps/Hz over a correlated rayleigh fading channel, where the concatenated channel matrix in slot-$t$ can be formulated as
\begin{align}
  \mathbf H(t) = \mathbf R_r^{1/2} \mathbf H_w(t) \mathbf R_t^{1/2}
\end{align}
where $\mathbf H_w(t)$ is normalized i.i.d.~rayleigh fading channel matrix while $\mathbf R_t$ and $\mathbf R_r$ are transmit and receive correlation matrices with $(i,j)$-th entry being $\tau_t^{|i-j|}$ and $\tau_r^{|i-j|}$~\cite{Loyka:01,Xiao:04}, respectively, where $\tau_t$ and $\tau_r$ are randomly chosen within $[0,1)$. Note that the users' channel vectors are the rows of $\mathbf H(t)$.

The parameters in the simulation are set as follows: maximum 500 gradient-descent iterations for the GMAT-MMSE, $\beta=0.01$.  The performance is averaged over 1000 channel realizations. Recall that the present channel coefficients (c.f.~$\mathbf D_{i}^{(k)}$, e.g., $h_{A1}(3)$ and $h_{B1}(3)$ for the 2-user case) are unknown for the transmitter and therefore are ignored for precoder design, while they should be taken into account at the receiver for MMSE receive filter design. Naturally, such a mismatch would result in performance degradation, but our proposed precoding methods are verified to be always effective thanks to the efficient trade-off between interference alignment and signal enhancement.


We show in Fig.~\ref{fig:subfig-k-2} for the 2-user case the sum rate comparison with MMSE receiver among GMAT-MMSE with the iteratively updated $\mathbf w_1$, $\mathbf w_2$, GMAT-DSINR with closed-form solutions in eq-(\ref{eq:MI-1-obj}-\ref{eq:MI-1-obj0}), and the original MAT algorithm with $\mathbf w_1=\mathbf h_{B}(1)$, $\mathbf w_2=\mathbf h_{A}(2)$, with the same power constraint $\|\mathbf w_1\|^2+\|\mathbf w_2\|^2 \le 2$ for all. 
In Fig.~\ref{fig:subfig-k-2}, the gap of sum rate between GMAT and MAT illustrates improvement of the GMAT-MMSE and GMAT-DSINR algorithms over the initial MAT concept, demonstrating the benefit of the trade-off between interference alignment and desired signal orthogonality enhancement. 
Compared with the original MAT algorithm, the two GMAT approaches have gained great improvement at finite SNR and possessed the same slope, which implies the same multiplexing gain, at high SNR. Interestingly, the closed-form solution performs as well as the iterative one, indicating the effectiveness of the mutual information approximation.

In Fig.~\ref{fig:subfig-k-3}, we present the similar performance comparison for the 3-user cases. The GMAT-MMSE solution updates order-2 message generation matrix $\mathbf W^{(2)}$ iteratively, while the original MAT algorithm set it according to eq-(\ref{eq:w-MAT}) and the GMAT-DSINR solution is obtained by optimizing eq-(\ref{eq:K-user-DSINR-O}) and eq-(\ref{eq:K-user-DSINR-C}). All these methods hold the same power allocation. With more transmit antennas and users, the same insights regarding the trade-off between signal orthogonality and interference alignment can be always obtained. It is interesting to note that, GMAT-DSINR performs as well as GMAT-MMSE, despite the distributed optimization.

\section{Conclusion}
We generalize the concept of precoding over a multi-user MISO channel with delayed CSIT for arbitrary number of users case, by proposing a precoder construction algorithm, which achieves the same DoF at infinite SNR yet reaches a useful trade-off between interference alignment and signal enhancement at finite SNR. Our proposed precoding concept lends itself to a variety of optimization methods, e.g., virtual MMSE and mutual information solutions, achieving good compromise between signal orthogonality and interference alignment.

\appendix \renewcommand\thesubsection{Appendix \Roman{subsection}}

\subsection{Gradient Descent Parameter for GMAT-MMSE}
Let $[\mathbf H_{ij}^l]_{m,n}=\mathbf e_m^H \mathbf H_{ij}^l \mathbf e_n$ be the $m$-th row and $n$-th column element of $\mathbf H_{ij}^l$. Particularly,
\begin{eqnarray}
[\mathbf H_{ij}^l]_{m,n}=\mathbf e_{m'}^H \mathbf W_j^l{(k)}  \mathbf e_n
\end{eqnarray}
when $m=\sum_{s=1}^{k-1} T_s + m'$ where $1 \le m' \le T_k$ and $1 \le n \le K$. Here, $\mathbf e_m$ is defined as the binary vector with only one `1' at $m$-th row.  By differentiating over $\mathbf W_j^l{(2)}$, we have
\begin{eqnarray}
&&\frac{\partial [\mathbf H_{ij}^l]_{m,n}}{\partial \mathbf W_j^{l~T}{(2)}} = \left(\frac{\partial [\mathbf H_{ij}^l]_{m,n}}{\partial \mathbf W_j^l{(2)}}\right)^T \\
&=& \left\{\begin{matrix} \mathbf 0 & if~m \le T_1 \\ \mathbf e_n \mathbf e_{m'}^H & if~T_1+1 \le m \le T_1+T_2 \\ \mathbf e_n \mathbf e_{m'}^H \prod_{t=2}^{k-1} \mathbf C^l(t) {\bf \Lambda}^l(t) & if~\sum_{s=1}^{k-1} T_s + 1 \le m \le \sum_{s=1}^{k} T_s~when~k \ge 3 \end{matrix}  \right. \\
&=& \mathbf e_n \mathbf e_{m}^H \mathbf Q^l
\end{eqnarray}
where
\begin{eqnarray}
\mathbf Q^l =  \begin{bmatrix} \mathbf 0_{T_1 \times K}  \\ \mathbf 0_{m_2^l \times K} \\ \mathbf I \\ \mathbf 0_{n_2^l \times K} \\ \vdots \\ \prod_{t=2}^{K-1} \mathbf C^l(t) {\bf \Lambda}^l(t)  \end{bmatrix}.
\end{eqnarray}
Note that we abuse vector $\mathbf e_{m'}$ with various dimensions $T_k$ according to the corresponding matrices $\mathbf W_j^l{(k)}$ for the sake of notational simplicity. Then, it follows that
\begin{eqnarray}
\frac{\partial [\mathbf H_{ij}^l]_{m,n}}{\partial [\mathbf W_j^{l~T}{(2)}]_{p,q}}
&=& \mathbf e_m^H \mathbf Q^l \mathbf e_p \mathbf e_q^H \mathbf e_n
\end{eqnarray}
where $1 \le p \le T_k$, $1 \le q \le K$, and we have
\begin{eqnarray}
\frac{\partial \mathbf H_{ij}^l}{\partial [\mathbf W_j^{l~T}{(2)}]_{p,q}} = \mathbf Q^l \mathbf e_p \mathbf e_q^H
\end{eqnarray}
Finally, according to the chain rule of matrix differentiation~\cite{Hjorungnes:07,Palomar:06}, we have
\begin{eqnarray}
\frac{\partial \left( J_i^l \right) }{\partial [\mathbf W_j^{l~T}{(2)}]_{p,q}} &=&  \text{Tr} \left( {\left(\frac{\partial J_i^l}{\partial \mathbf H_{ij}^l}\right)}^{T} \frac{\partial \mathbf H_{ij}^l}{\partial \mathbf [\mathbf W_j^{l~T}{(2)}]_{p,q}} \right) = \text{Tr} \left( \mathbf e_q^H  {\left(\frac{\partial J_i^l}{\partial \mathbf H_{ij}^l}\right)}^T \mathbf Q^l \mathbf e_p \right)
\end{eqnarray}
So, for the $K$-user case, the Gaussian descent parameter can be calculated by
\begin{eqnarray}
\frac{\partial \left(J \right)}{\partial \mathbf W_j^l{(2)}} = \sum_{i=1}^K \frac{\partial \left( J_i^l \right)}{\partial \mathbf W_j^l{(2)}} &=& \sum_{i=1}^K  {\left(\frac{\partial J_i^l}{\partial \mathbf H_{ij}^l}\right)}^{T} \mathbf Q^l
\end{eqnarray}
where
\begin{eqnarray} \label{eq:differentiation-start}
{\left(\frac{\partial J_i^l}{\partial \mathbf H_{ii}^l}\right)}^{T} = f \left( \sqrt{\rho}\mathbf H_{ii}^l, \rho \sum_{l=1}^L \sum_{j=1, j \ne i}^K \mathbf H_{ij}^l\mathbf H_{ij}^{l~H}+ \mathbf I \right)\\
{\left(\frac{\partial J_i^l}{\partial \mathbf H_{ij}^l}\right)}^{T} = g \left( \sqrt{\rho}\mathbf H_{ij}^l, \rho \sum_{l=1}^L \sum_{k=1, k \ne j}^K \mathbf H_{ik}^l\mathbf H_{ik}^{l~H}+\mathbf I \right)
\end{eqnarray}
where
\begin{eqnarray}
f(\mathbf A, \mathbf B)&=&-\mathbf A^H \left(\mathbf A \mathbf A^H + \mathbf B \right)^{-1} \mathbf B \left(\mathbf A \mathbf A^H + \mathbf B \right)^{-1}\\
g(\mathbf A, \mathbf B)&=&\mathbf A^H \left(\mathbf A \mathbf A^H + \mathbf B \right)^{-1} (\mathbf B-\mathbf I) \left(\mathbf A \mathbf A^H + \mathbf B \right)^{-1}
\end{eqnarray}

\vspace{-20pt}
\section*{acknowledgements}
Fruitful discussions with Sheng Yang, Mari Kobayashi and Paul de Kerret are thankfully acknowledged.

\nocite{*}
\bibliographystyle{IEEE}

\begin{figure}
\begin{center}
\includegraphics[width=0.6\columnwidth]{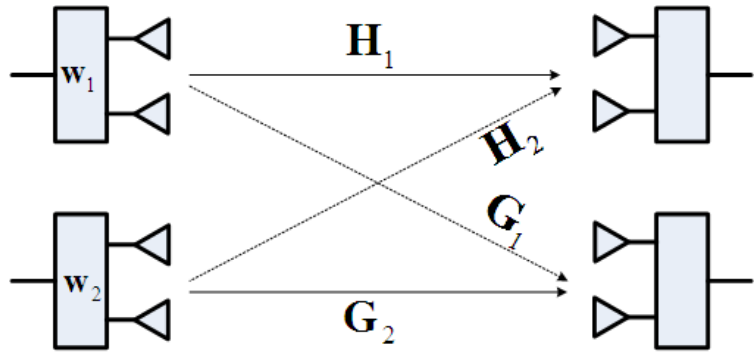}
\caption{Interpretation as MIMO Interference Channel.}
\end{center}
\end{figure}

\newpage
\begin{figure}
\centering
\includegraphics[width=0.8\columnwidth]{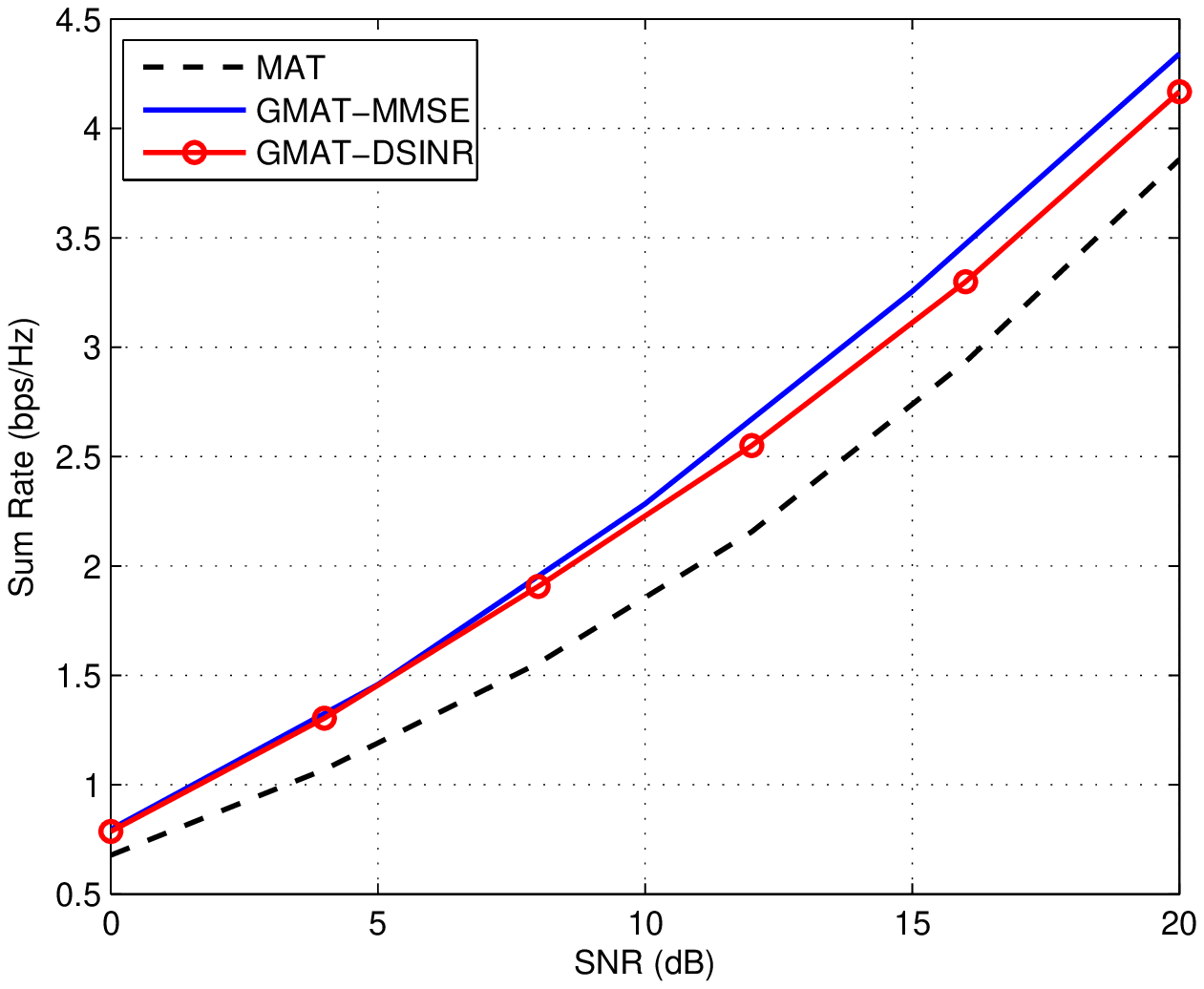}
\caption{Sum rate vs. SNR for the 2-user case.}
\label{fig:subfig-k-2}
\end{figure}

\begin{figure}
\centering
\includegraphics[width=0.8\columnwidth]{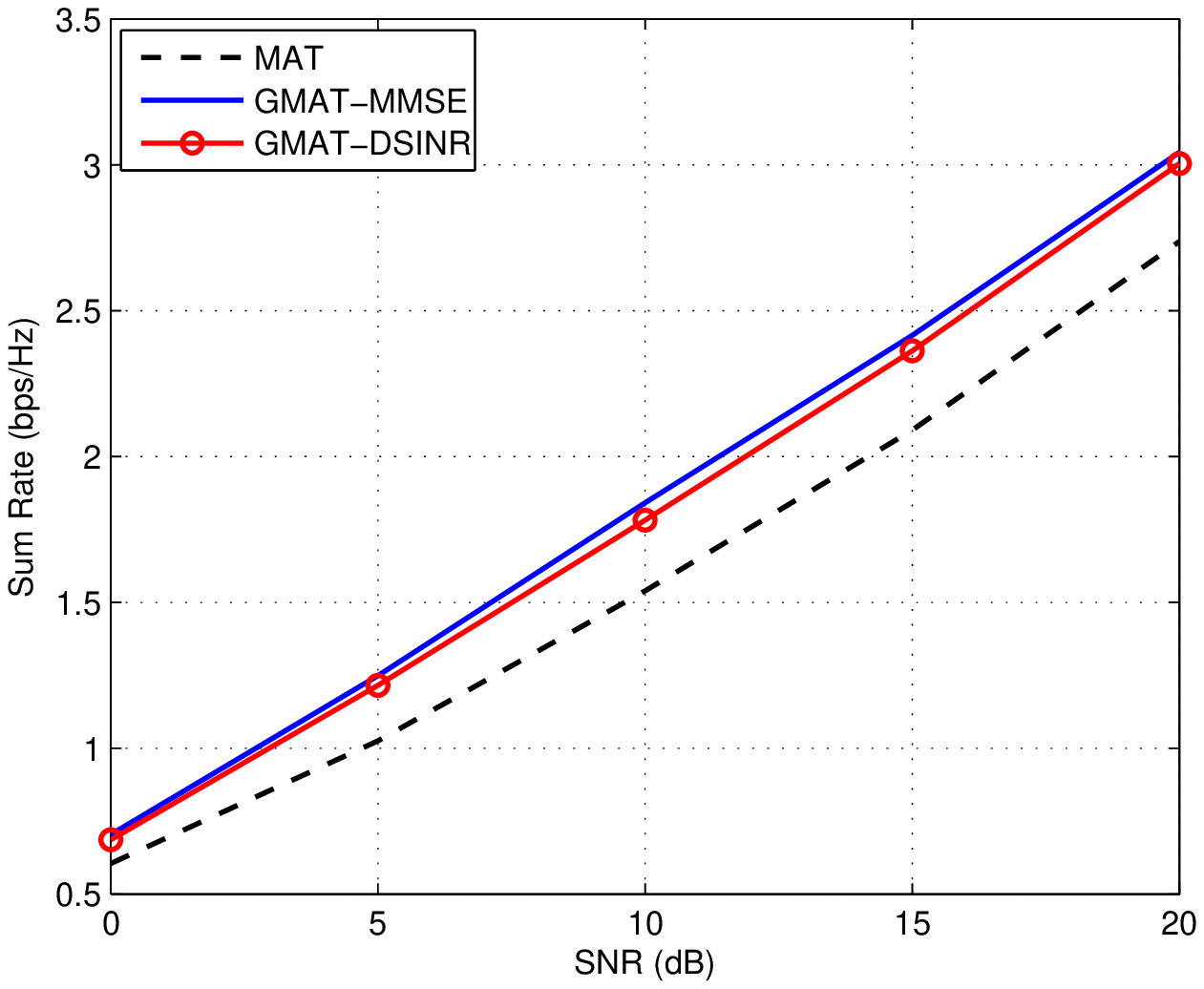}
\caption{Sum rate vs. SNR for the 3-user case.}
\label{fig:subfig-k-3}
\end{figure}

}
\end{document}